\documentclass[sigconf,natbib=true]{acmart}
\usepackage{algorithm,algorithmic}

\usepackage{threeparttable}
\usepackage{booktabs} 
\usepackage{multirow}
\usepackage{graphicx}

\usepackage{amsmath}
\usepackage{cases}
\usepackage{booktabs}

\usepackage{subfigure} 
\usepackage{enumitem}
\usepackage{caption}

\allowdisplaybreaks[4]

\usepackage{color}
\usepackage{tabularx}

\acmConference[]{}{}{}
\begin{document}

\title{FedHAP: Federated Hashing with Global Prototypes for Cross-silo Retrieval}
\author{Meilin Yang, Jian Xu, Yang Liu, Wenbo Ding}
\affiliation{%
	\institution{Tsinghua-Berkeley Shenzhen Institute (TBSI), Tsinghua University}
	\city{Shenzhen}
	\country{China}
}
\affiliation{%
	\institution{Institute for AI Industry Research, Tsinghua University}
	\city{Beijing}
	\country{China}
}


\begin{abstract}
Deep hashing has been widely applied in large-scale data retrieval due to its superior retrieval efficiency and low storage cost. However, data are often scattered in data silos with privacy concerns, so performing centralized data storage and retrieval is not always possible. Leveraging the concept of federated learning (FL) to perform deep hashing is a recent research trend. However, existing frameworks mostly rely on the aggregation of the local deep hashing models, which are trained by performing similarity learning with local skewed data only. Therefore, they cannot work well for non-IID clients in a real federated environment. To overcome these challenges, we propose a novel federated hashing framework that enables participating clients to jointly train the shared deep hashing model by leveraging the prototypical hash codes for each class. Globally, the transmission of global prototypes with only one prototypical hash code per class will minimize the impact of communication cost and privacy risk. Locally, the use of global prototypes are maximized by jointly training a discriminator network and the local hashing network. Extensive experiments on benchmark datasets are conducted to demonstrate that our method can significantly improve the performance of the deep hashing model in the federated environments with non-IID data distributions.

\end{abstract}

\maketitle

\section{Introduction}
With the explosive increase of data generated from different institutions, achieving fast and storage-saving information retrieval across multiple institutions has attracted much attention in recent years. Deep hashing is a widely used method that aims to reduce storage cost and improve retrieval efficiency by encoding data points into non-invertible and compact binary hash codes with deep neural networks (DNNs) \cite{he2016deep,krizhevsky2012imagenet}. Most existing deep hashing methods assume that data storage is centralized. For example, TDHPPIR \cite{zhang2020tdhppir} is an efficient privacy-preserving image retrieval method based on deep hashing, in which data owners upload the encrypted data set to the central cloud server, and provide data retrieval services via the indexes established from the database. However, centralized storage of private client data is not always feasible due to space limitations, increasing privacy concerns and tough data protection regulations such as GDPR \footnote{https://gdpr-info.eu}. Therefore, it is increasingly desirable to learn to hash over distributed data with privacy protection.

In recent years, federated learning (FL) \cite{konevcny2016federated} has emerged as a promising paradigm for collaborative learning with privacy preservation. In the original FL framework, the popular FedAvg algorithm is proposed. In FedAvg, the selected clients first locally perform multiple training epochs by stochastic gradient descent (SGD), and then transmit their model updates to a central server, where the model updates are aggregated to obtain a new global model. Previous works \cite{xu2020federated,zong2021fedcmr} have explored the combination of deep hashing with FL and demonstrated their effectiveness, but these works simply rely on model aggregation to achieve global hash learning, which does not sufficiently address the non-IID\footnote{Data distributions across clients are not identical.} nature of data in federated environments. For example, in a patient hashing problem across multiple hospitals, the distribution of patients from an oncology hospital and a psychiatric hospital can be quite different, so can the data quantity. Since the core of deep hashing models is data similarity learning, skewed and highly imbalanced local data distributions can result in biased local models that cannot be sufficiently corrected by the FedAvg algorithm.

To tackle the above issues, in this paper, we introduce a \textbf{fed}erated deep \textbf{ha}shing method with global \textbf{p}rototypes (FedHAP) for cross-silo retrieval. In the FedHAP framework, each client in the federation can jointly train the hashing model using its own local data and global prototypical hash codes of each class to guide the local training. Specifically, the global prototypes are generated in the server by aggregating the class-averaged hash codes from clients. Then the prototypes, together with the global model, are broadcast to clients for local training. To better utilize the global prototypes locally, we not only design similarity learning algorithms with supervision from the global hash codes, but also creatively design a discriminator network to ensure the distribution consistency between the locally generated binary hash codes and the global ones. By this way, we maximize the usage of global prototypes to enhance the local training of each client without exchanging sample-level information, thereby significantly improving the performance of the federated hashing model while preserving the privacy of local data. {During the retrieval process, the hash code of query generated by the trained hashing model will be sent to each client, and the best matching data will be retrieved by finding the data with the shortest similarity distance.}

\begin{figure}[htbp]
\centering
\subfigure[DPSH(FedAvg)]{
\includegraphics[width=3.8cm]{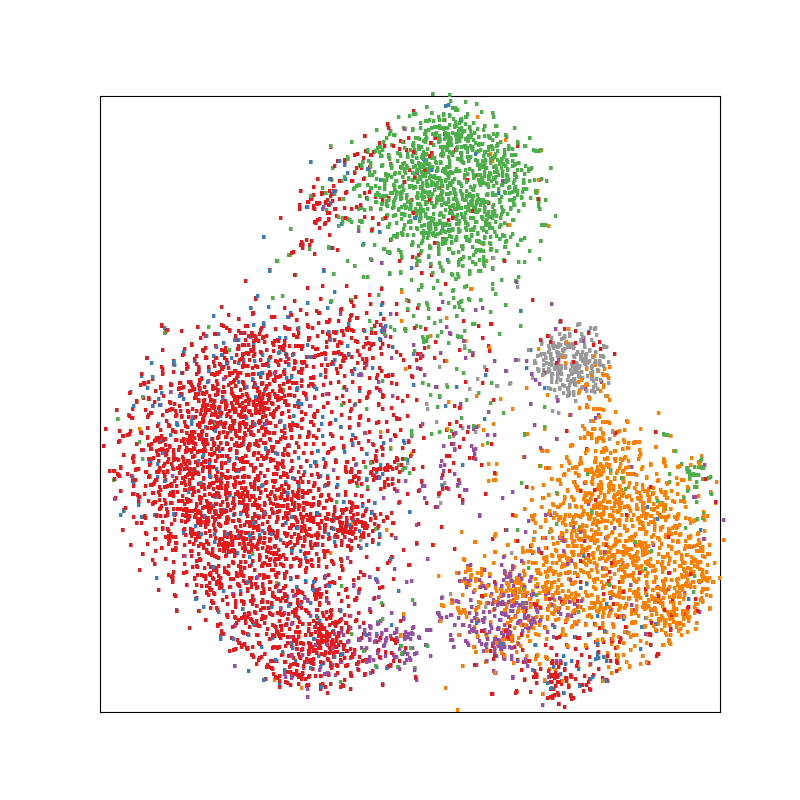}
}
\quad
\subfigure[FedHAP]{
\includegraphics[width=3.8cm]{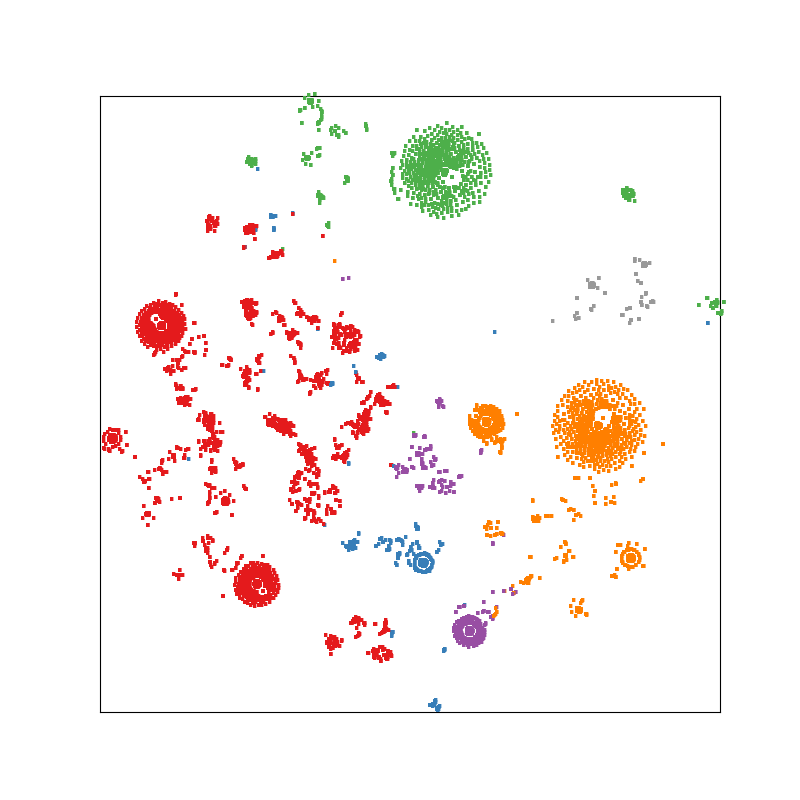}
}
\caption{Visualization of hash codes generated by the two deep hash learning methods, trained on 10000 training data points of NUS-WIDE for different classes. (For ease of visualization, we sample six categories)}
\label{fig:tsne}
\vspace{-3.4ex}
\end{figure}

As an example for demonstrating the efficacy of our proposed FedHAP, we compare the hash codes generated by our approach and the naive FedAvg approach as in Fig.~\ref{fig:tsne}, where each data point is visualized using t-SNE \cite{van2008visualizing}. We observe that the hash codes learned by FedHAP exhibit favorable intra-class compactness and inter-class separability compared with the baseline that adopts deep supervised hashing using pairwise labels (DPSH) \cite{li2015feature} with FedAvg \cite{li2020federated} for hashing. It is worth noting that our approach works especially better for under-represented classes with fewer samples, as demonstrated by the clear discrimination of the two categories in blue and purple. 


The major contributions of this paper are summarized as follows:
\begin{itemize}[leftmargin=*]
\item We present a novel federated supervised hashing method named FedHAP for efficient and effective cross-silo retrieval. This method integrates hashing learning with federated learning and takes advantage of the global prototypes, to enhance the performance of the hashing model with minimal impact on privacy.

\item The global prototypes are leveraged in both hashing learning and our introduced adversarial learning to enforce the semantic consistency between local hash codes and global prototypes, which can align the local learned distributions of hash codes and thus facilitate the global model aggregation.

\item Experimental results on three benchmark datasets demonstrate that our approach outperforms existing methods and can achieve significantly improved mAPs in both IID and non-IID scenarios. Furthermore, we verify the efficacy of each component in our proposed method by ablation experiments.
\end{itemize}


\section{RELATED WORK}
\subsection{Deep Hashing}
Hashing functions have been attractive due to their irreversible nature, which can map sensitive data into compact binary codes. Analytics has proven that the learned binary hash codes can lead to less memory consumption and short query time \cite{chi2017hashing}. Existing hashing methods can be generally organized into two categories: unsupervised hashing and supervised hashing. Unsupervised hashing methods \cite{weiss2008spectral,gong2012iterative,kong2012isotropic,liu2014discrete} learn hashing functions that map input data points into binary codes by exploiting the similarity distances of samples. Supervised hashing methods \cite{erin2015deep,liu2012supervised,rastegari2012attribute,cao2018deep,li2015feature,su2018greedy,yuan2020central} aim to further exploit available supervised information (such as labels or the semantic affinities of training data) to improve performance. In recent years, supervised hashing has attracted more attention as it can achieve better accuracy than unsupervised hashing. Deep Convolutional Neural Network based hashing methods \cite{erin2015deep,lai2015simultaneous,xia2014supervised,zhang2015bit,zhao2015deep} are proposed to learn data representations in binary codes that preserve the locality and similarity of data samples. By coupling data feature extraction and binary code learning, these methods have been shown to greatly improve retrieval accuracy.

\vspace{-3.6ex}




\subsection{Federated Learning}

FL was first proposed by Google in 2016 \cite{konevcny2016federated}, and has emerged as a paradigm for distributed training of machine learning models without direct data-sharing. It enables multiple clients to collectively learn a model under the coordination of a central server while keeping the data decentralized and protecting the data privacy of each client. FedAvg \cite{li2020federated} is a popular optimization method in federated learning but relies on the premise that each local solver is a copy of the same stochastic process (due to the IID assumption). FedProx \cite{mcmahan2017communication} is presented as a generalization and re-parametrization of FedAvg, and it achieves stable and accurate convergence behavior compared to FedAvg in highly heterogeneous settings. To address the non-IID nature of FL, various algorithms have been proposed since then \cite{lin2020ensemble,li2021model,wang2020tackling,hsu2019measuring,mu2021fedproc,wang2020federated}. FedProc \cite{mu2021fedproc} designs a local network architecture and a contrastive loss to regulate the training of local models with the class-wise logits transmission. However, unprocessed raw intermediate logits may cause leakage of the original data and local data distribution. Knowledge Distillation based federated frameworks \cite{lin2020ensemble,lopes2017data} have recently emerged to tackle the non-IID issue by refining the server model using aggregated knowledge from heterogeneous users, which may need a proxy dataset or require the client to provide the server with the label distribution.

\vspace{-3ex}

\subsection{Federated Hashing}
Recently, the framework of federated learning has been applied to hashing for various tasks \cite{xu2020federated,zong2021fedcmr}. For example, Federated Patient Hashing (FPH) \cite{xu2020federated} has been proposed to collaboratively train a patient information retrieval model stored in a shared memory while keeping all the patient-level information in local clients. Federated Cross-Modal Retrieval (FedCMR) \cite{zong2021fedcmr} is the first attempt to combine federated learning with cross-modal retrieval. While the abovementioned methods have certainly proved the feasibility of federated hashing to some extent, they did not exploit the relationship between the global hashing model and the local hashing model in the FL framework, nor did they sufficiently address the prominent non-IID problem in the federated environments, such as label distribution skewness. Different from these methods, our approach design a mechanism to extract the statistics of class-wise global prototypes, and symbolic operations are also performed on the class-wise global prototypes, which greatly avoids the leakage of data and label distribution. The class-wise global prototypes will participate the local training process and alleviate the model drift issue. Furthermore, illuminated by adversarial learning, we design a discriminator network to further bridge the global and local generation of hash codes. 




\begin{figure*}[h]
  \centering
  \setlength{\abovecaptionskip}{-1cm}
  \includegraphics[width=1.1\textwidth]{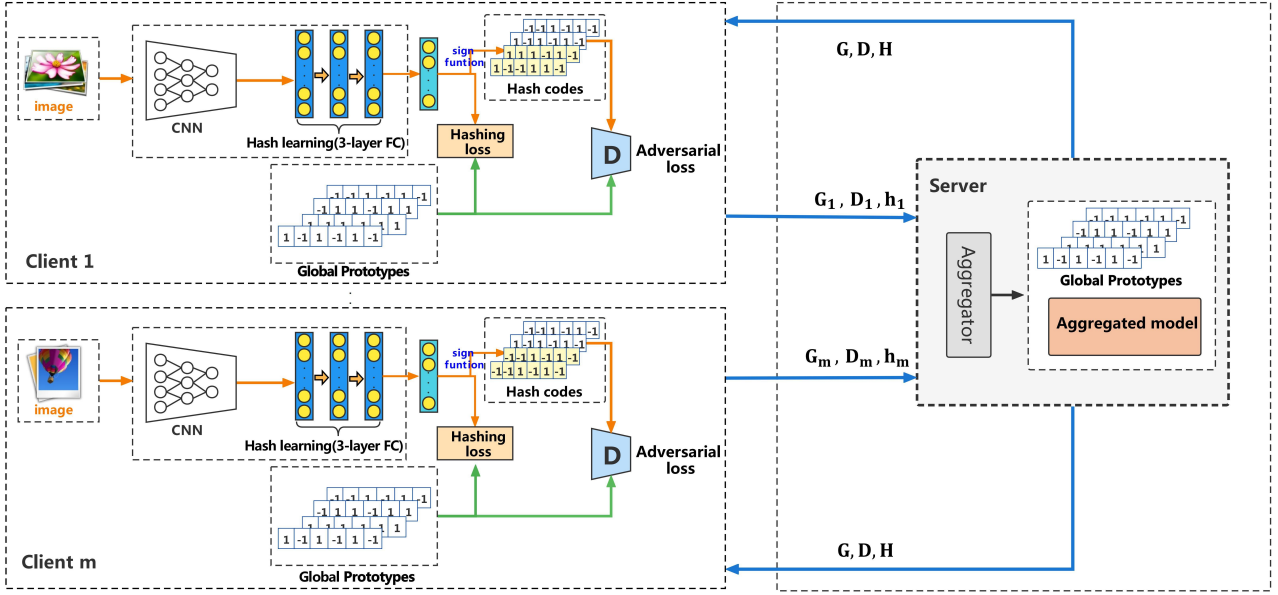}
  \caption{The framework of our proposed FedHAP.}
  \label{fig:framework}
\end{figure*}

\section{THE PROPOSED METHOD}
\label{sec:method}
In this section, we first present the problem definitions. Then, we introduce the details of our FedHAP approach and provide analysis on each of the design components. 

\subsection{Problem Formulation}
Without losing generality, we assume there are $m$ clients whose private data are denoted as $\mathcal{D}_i =\{\mathbf{x}_j\}_{j=1}^{{n}_{i}}$, where $\mathbf{x}_j \in \mathbb{R}^{1 \times d}$ is the input features, ${n}_{i}$ is the number of data in client $i$, and $d$ is the dimension of features. All clients collaboratively train a hashing model $G$ with parameters $\theta^G$, which maps the input features $\mathbf{x}_j$ into output $\mathbf{b}_j \in \mathbb{R}^{1 \times K}$, where $K$ is the number of bits in Hamming space. 
The hash code of $\mathbf{x}_j$ is denoted as ${\mathbf{h}_j}$ and can be gained by ${\mathbf{h}_j}=sign({\mathbf{b}_j})$, where $sign$ denotes the element-wise sign function. Let $\mathcal{H}_i = \{\mathbf{h}_j\}_{j=1}^{{n}_{i}}$, the problem to tackle here is for $m$ clients to collaboratively train a deep hashing model without exposing their data $\mathcal{D}_i$, so that the similarities among all data sample pairs are preserved as follows:
\begin{equation}
\min_{{\theta^G}} \sum_{i=1}^m \mathcal{L}_{hash}^{i}(\mathcal{D}_i, \mathcal{H}_i).
\label{eq:formulate loss}
\end{equation}
where $\mathcal{L}_{hash}^{i}$ denotes the hashing loss which we will explain next.
 
  




\subsection{Learning the Hashing Model}
The deep hashing model $G$ is comprised of a feature learning module and a hash learning module. The feature learning  module is a deep convolutional neural network to extract representations from the input data,  
which is then fed into the hash learning module. The hash learning module consists of multiple fully connected layers to obtain $\mathbf{b}_j$. 



\vspace{1ex}
\noindent \textbf{Similarity preserving loss ($\mathcal{L}_{tl}$).} To ensure that the hash codes of similar data pairs are pulled together and the codes of dissimilar data pairs are pushed away from each other, we choose a widely used triplet ranking loss\cite{weinberger2006distance} to preserve the similarity structure between data pairs. In order to explain the triplet loss briefly, we suppose there is a set of triplets $\left (\mathbf{b}_{j}, \mathbf{b}_{j^{+}}, \mathbf{b}_{j^{-}} \right )$. $\mathbf{b}_{j^{+}}$ is a positive pair of $\mathbf{b}_{j}$, which indicates $\mathbf{b}_{j}$ and $\mathbf{b}_{j^{+}}$ have the same class. $\mathbf{b}_{j^{-}}$ is a negative pair of $\mathbf{b}_{j}$, which indicates $\mathbf{b}_{j}$ and $\mathbf{b}_{j^{-}}$ are of the different classes. The similarity between the codes can be evaluated using a general distance metric ${d\left(\cdot,\cdot \right)}$, e.g., cosine distance. For example, ${d}\left ( \mathbf{b}_{j},\mathbf{b}_{j^{+}} \right )$ computes the dissimilarity between the sample pairs. The triplet loss mentioned above can be formulated as follows:
\begin{equation}
\setlength{\abovedisplayskip}{3pt}
\setlength{\belowdisplayskip}{3pt}
\mathcal{L}_{tl^{local}}^{i}=\sum_{j=1}^{{n}_{i}}max\left ({d}\left ( \mathbf{b}_{j},\mathbf{b}_{j^{+}} \right ) -{d}\left ( \mathbf{b}_{j},\mathbf{b}_{j^{-}} \right )+a,0 \right ).
\label{eq:local sim loss}
\end{equation}
where $a$ is the margin parameter. Note this loss can be only computed locally for each client $i$ as denoted by the $local$ mark on $\mathcal{L}_{tl^{local}}^{i}$, as clients are not supposed to exchange raw data. To enhance hash learning by leveraging other clients' prototypes, we further consider a novel global triplet loss between local data and global prototypes of the hash codes for each class, denoted as $\mathcal{L}_{tl^{global}}^{i}$, see Eq.~\eqref{eq:global sim loss}. Here, the global prototypes collectively are denoted as $\hat{\mathcal{H}}=\{ \mathbf{\hat{h}}_{c}\}_{c=1}^{C}$, where $C$ is the number of classes and $\mathbf{\hat{h}}_c$ is the prototypical hash code for class $c$. In Eq.~ \eqref{eq:global sim loss}, $\mathbf{\hat{h}}_j^{+}$ denotes the global hash code that is of the same class as $\mathbf{b}_j$, and $\mathbf{\hat{h}}_j^{-}$ denotes the global hash code that is of a different class from $\mathbf{b}_j$. We will discuss how to generate $\hat{\mathcal{H}}$ in the following. 
\begin{equation}
\mathcal{L}_{tl^{global}}^{i}=\sum_{j=1}^{{n}_{i}}
        max\left (d\left ( \mathbf{b}_{j}, \mathbf{\hat{h}}_{j^{+}}  \right ) -{d}\left ( \mathbf{b}_{j},\mathbf{\hat{h}}_{j^{-}}  \right )+a,0 \right ).
\label{eq:global sim loss}
\end{equation}
\begin{equation}
\mathcal{L}_{tl}^i=\mathcal{L}_{tl^{local}}^i+\mathcal{L}_{tl^{global}}^{i}.
\label{eq:local-global sim loss}
\end{equation}

\begin{figure}
  \centering
  \includegraphics[width=0.95\columnwidth]{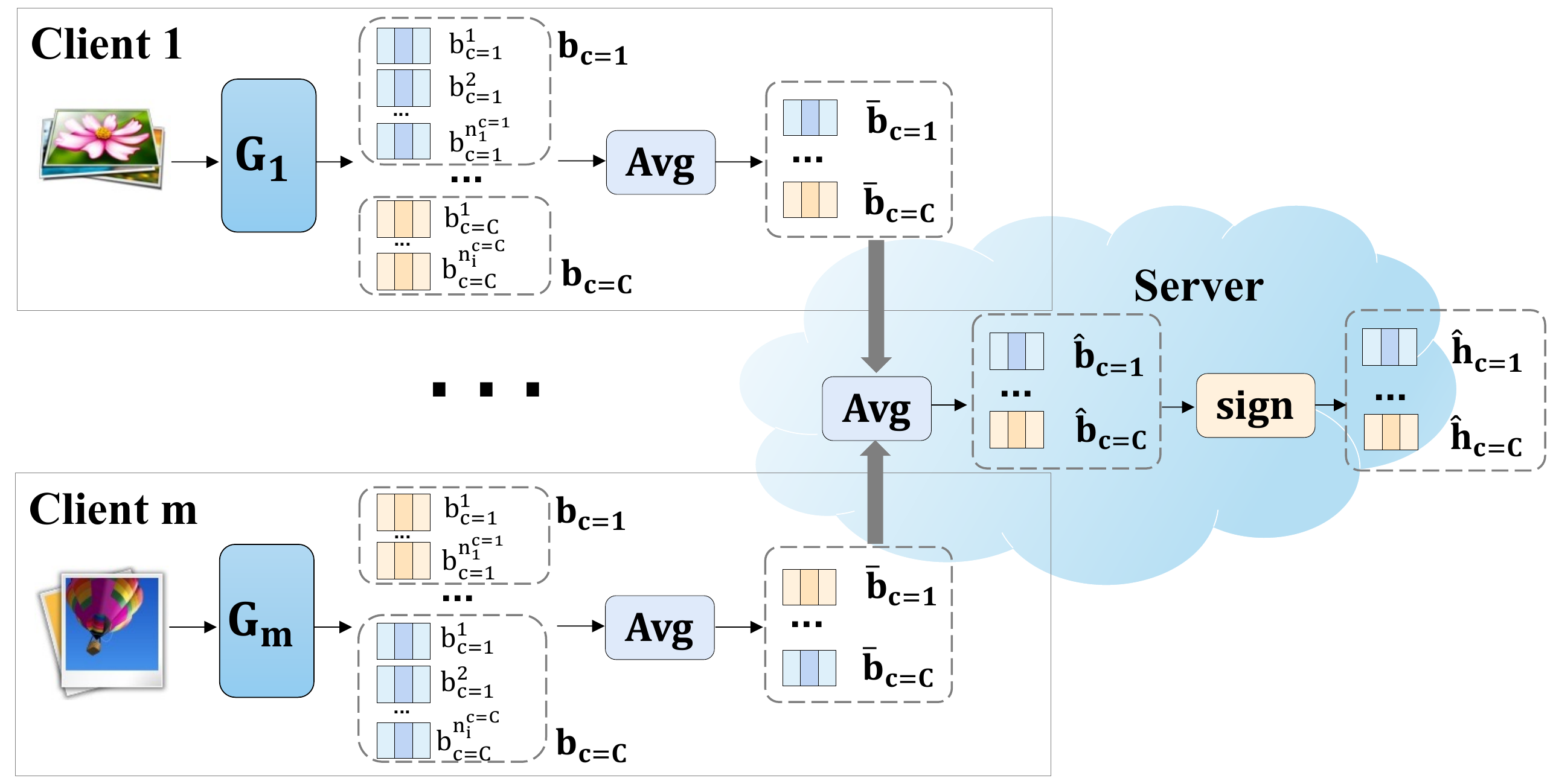}
  \caption{The generation of the global prototypes.}
  \label{fig:global}
  \vspace{-2ex}
\end{figure}

At each round, the global hash codes $\hat{\mathcal{H}}$ will be computed by $\hat{\mathcal{H}}=sign\left ( \hat{\mathcal{B}}\right )$, and $\hat{\mathcal{B}}=\{ \mathbf{\hat{b}}_{c}\}_{c=1}^{C}$ is obtained by aggregating all the class-level vectors from clients as illustrated in Fig.\ref{fig:global}. First, each client aggregates their class prototypical hash codes by Eq.~ \eqref{eq:b}. 
 \begin{equation}
\mathbf{\bar{b}}_{i,c}=\frac{\sum_{j=1}^{{n}_{i,c}}\mathbf{b}_{j,c}}{{n}_{i,c}}.
\label{eq:b}
\end{equation}
where ${n}_{i,c}$ indicates the number of data samples of class $c$ in client $i$ and $\mathbf{b}_{j,c}$ represents the ouput feature vector of the $j$-th data of class $c$.
Next, clients send their class-level hash codes to the server which performs the aggregation:
\begin{equation}
\mathbf{\hat{h}}_{c}=sign\left (\mathbf{\hat{b}}_{c}\right )=sign\left ( \frac{1}{m}\sum_{i=1}^{m}\mathbf{\bar{b}}_{i,c}\right ).
\label{eq:h}
\end{equation}






\vspace{2ex}
\noindent \textbf{Quantization loss ($\mathcal{L}_{quan}$).}
The binary constraints on $\mathbf{b}_{j}$ require thresholding the network outputs (e.g. with a sign function), which will make it difficult to train the network with backpropagation. To simplify the optimization process during the hashing learning, the common way is to solve the relaxed problem through dropping the sign function, which will introduce the non-negligible quantization loss. To overcome this problem, we introduce the approximation loss for the learned hash codes in Eq.~\eqref{eq:quant loss},
\begin{equation}
\mathcal{L}_{quan}^i=\sum_{j=1}^{{n}_{i}}\left \|\mathbf{b}_{j}^{}-sign\left ( \mathbf{b}_{j} \right )  \right \|^{2}.
\label{eq:quant loss}
\end{equation}

\begin{figure}[ht]
  \centering
  \includegraphics[width=0.43\textwidth,height=0.19\textwidth ]{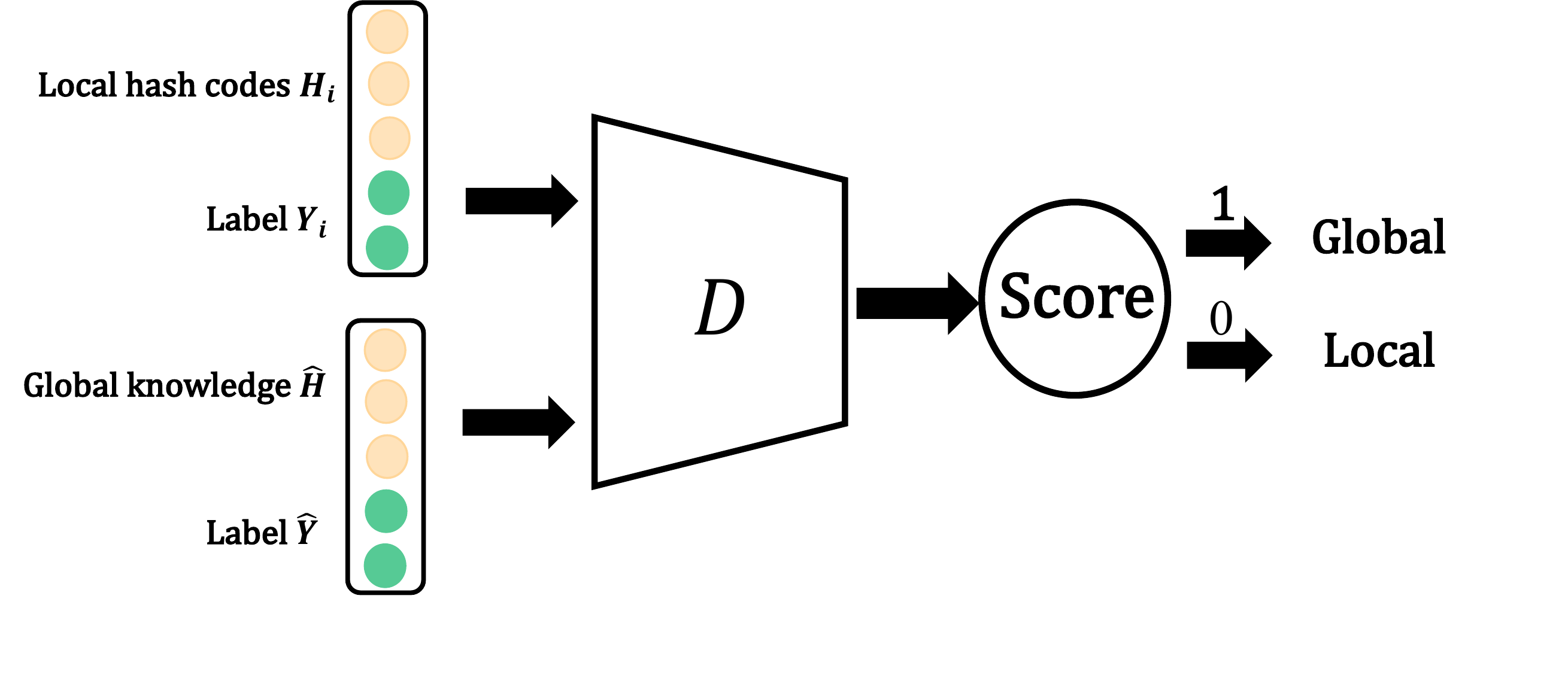}
  \vspace{-3ex}
  \caption{The workflow of the proposed discriminator $D$.}
  \label{fig:dis}
  \vspace{-2ex}
\end{figure}

\vspace{1ex}
\noindent \textbf{Adversarial loss ($\mathcal{L}_{d}$).} In the federated scenario with non-IID distributions, the consistency of the generated hash codes from different clients cannot be guaranteed. In order to preserve the consistency of local and global distributions of hash codes, we further introduce a local discriminator network $D_i$ for each client with trainable parameters $\theta^{D_i}$. The discriminator network is initially used in adversarial learning to identify whether the data come from a real dataset or a neural network \cite{goodfellow2014generative} and its output is the probability that the input data come from the real dataset. In this paper, we treat the global hash codes $\mathcal{\hat{H}} = \{\mathbf{\hat{h}}_{c}\}_{c=1}^{C}$ as the real dataset and the hash codes $\mathcal{H}_i = \{\mathbf{h}_j\}_{j=1}^{{n}_{i}}$ generated by the local hashing model as the latter. We utilize the local labels $Y_{i}=\{\mathbf{y}_{j}\}_{j=1}^{{n}_{i}}$ and global labels $\hat{Y}=\{\mathbf{\hat{y}}_{c}\}_{c=1}^{C}$ as constraints on $\mathcal{H}_i$ and $\mathcal{\hat{H}}$ to realize the discrimination of hash codes of a specific class. Specifically, we use the one-hot vector of the class label as extra information and concatenate it with $\mathcal{H}_i/\mathcal{\hat{H}}$ together as the input vectors of $D$, and the output is the probability score (between 0 and 1) that the input data come from the global prototypes as shown in Fig.\ref{fig:dis}. Specifically, the score approaches to “1” when the input vector is classified as global prototypes, and vice versa. We define the adversarial loss as $\mathcal{L}_{d}$, which is a cross-entropy loss. The adversarial loss $\mathcal{L}_{d}^{i}$ of client $i$ can be written as follows:
\vspace{-1cm}
\begin{equation}
\setlength{\abovedisplayskip}{30pt}
\mathcal{L}_{d}^i=-\left ( \frac{1}{{n}_{i}}\sum_{j=1}^{{n}_{i}} \left ( 1-log( D_i\left (\mathbf{h}_{j}|\mathbf{y}_{j} \right )\right )+\frac{1}{C}\sum_{c=1}^{C} \left ( log( D_i\left ( \mathbf{\hat{h}}_{c}|\mathbf{\hat{y}}_{c} \right )\right ) \right).
\label{eq:adv loss}
\end{equation}
where the first term in the above equation is the cross-entropy loss for the local dataset, followed by the cross-entropy loss for the global prototypes.

\noindent \textbf{{Overall local objective.}}
The overall local loss function $\mathcal{L}_{hash}^{i}$ can be obtained by combining Eq.~\eqref{eq:local-global sim loss}, Eq.~\eqref{eq:quant loss} and Eq.~\eqref{eq:adv loss}, formulated as follows:
for the global prototypes.
\begin{equation}
\mathcal{L}_{hash}^i =\mathcal{L}_{tl}^i +\mu*\mathcal{L}_{quan}^i +\lambda*\mathcal{L}_{d}^i.
\label{eq:hash loss}
\end{equation}
where $\mu$ and $\lambda$ are two penalty parameters to balance different loss components.

\vspace{-2ex}
\subsection{FedHAP Framework and Algorithm}
For all clients to learn the above hashing model collaboratively, we propose FedHAP, which is shown in Fig.~\ref{fig:framework} and algorithm \ref{alg:alg}. The general framework mainly consists of a central server and $m$ clients. First, each client $i$ trains the deep hashing model and discriminator network with its local data and global prototypes, uploading the local updated model parameters $\theta^{G_{i}}$, $\theta^{D_{i}}$ and the locally generated prototypical codes $\mathbf{\bar{b}}_{i}=\{\mathbf{\bar{b}}_{i,c}\}_{c=1}^C$ to the central server. The central server is responsible for coordinating clients in the model training process by aggregating $\theta^{G_i}$, $\theta^{D_{i}}$ and $\mathbf{\bar{b}}_i$ received from clients and then delivering the aggregated models $\theta^{G_{i}}$, $\theta^{D_{i}}$ and $\hat{\mathcal{H}}$ to them for the next training round.

\vspace{1ex}
\noindent \textbf{Local training procedure.} During each local training step, the original input of data will be converted into low-dimensional features by the convolutional neural network and the hash learning module, which are then used to compute the similarity preserving loss $\mathcal{L}_{tl}^i$ with the guidance of global prototypes. Next, the feature embedding will be converted into binary hash codes using the sign function with quantization loss $\mathcal{L}_{quan}^i$. Furthermore, the local and global hash codes with their semantic labels will be simultaneously fed into the discriminator network to generate the corresponding adversarial loss $\mathcal{L}_{d}^i$. It is worth noting that the parameters of the hashing network and discriminator network rely on each other in the training process, and both of the two training phases will update all model parameters.

\begin{minipage}{.95\linewidth}
		\label{alg:alg}
		\renewcommand{\algorithmicrequire}{\textbf{Input:}}
		\renewcommand{\algorithmicensure}{\textbf{Initialize:}}
		\begin{algorithm}[H]
		    \caption{FedHAP}
			\begin{algorithmic}
			
				\REQUIRE Image set $\mathcal{X}$, Number of clients $m$, Hashing model $G$, Discriminator network $D$, communication rounds $T$, Local training epochs $E$.
				\ENSURE Initialize $\theta_{0}^{G}, \theta_{0}^{D}$, $\hat{\mathcal{H}^{0}}$
				\FOR {$t= 0$ to $T-1$}
				\STATE Server broadcasts global model parameters of  ${\theta_{t}^{{G}}}, {\theta_{t}^{D}}$ and global prototypes $  \hat{\mathcal{H}^{t}} $ to each client $i$.
				\FOR {each client $i$ \textbf{in parallel}}
				\FOR {$e= 1$ to $E$}
				\STATE Start training $D_{i}\left ( G_{i}\left(x \right), \theta_{t}^{D_{i}} \right )$.
				\STATE Calculate adversarial loss $\mathcal{L}_{d}^i$ with Eq.~\eqref{eq:adv loss}
				\STATE Update $\theta_{t}^{D_{i}}$ using back propagation
				\STATE Start training $G_{i}\left ( x, \theta_{t}^{G_{i}}  \right ) $
				\STATE Calculate adversarial loss $\mathcal{L}_{d}^i$, cosine triplet loss $\mathcal{L}_{tl}^i$, quantization loss $\mathcal{L}_{quan}^i$
				\STATE $\mathcal{L}_{hash}^i =\mathcal{L}_{tl}^i +\mu*\mathcal{L}_{quan}^i +\lambda*\mathcal{L}_{d}^i$
				\STATE Update $\theta_{t}^{G_{i}}$ using back propagation
				\ENDFOR
				\STATE Send local model parameters $\theta_{t}^{G_{i}}$, $\theta_{t}^{D_{i}}$, and local prototypes $\mathbf{\bar{b}}_{i}$ to the central server
				\ENDFOR
				\vspace{1ex}
				\STATE Server executes:
				\STATE \quad Update global hashing model ${\theta_{t+1}^{G}}\leftarrow \frac{1}{m} \sum_{i =1}^{m}\theta_{t}^{G_{i}}$
				
				\vspace{1ex}
				\STATE \quad Update global discriminator ${\theta_{t+1}^{D}} \leftarrow \frac{1}{m} \sum_{i =1}^{m} \theta_{t}^{D_{i}}$
				
				\vspace{1ex}
				\STATE \quad Update global prototypes $ \hat{\mathcal{H}}^{t+1}$
				
				\ENDFOR
			\end{algorithmic}
		\end{algorithm}
		\vspace{1ex}
	\end{minipage}

\noindent \textbf{Privacy and Communication concerns.}
Our proposed FedHAP requires the transmission of per-class prototypes between clients and server. However, this does not raise higher communication costs and privacy risks since the prototypes represent only an averaged statistic over all local data of low-dimensional feature representations, such as a 12-bit vector containing only -1 or 1. Similar methods have also been investigated in the literature \cite{wu2021hierarchical, mu2021fedproc}, where some statistics or prior knowledge under the premise of privacy protection are transmitted to facilitate the learning of federated models. 



\vspace{-1ex}
\section{Experiments}
In this section, we conduct extensive experiments to verify the effectiveness of our proposed approach and compare it with other state-of-the-art methods in federated environments, considering both IID and non-IID scenarios. We evaluate all methods on three benchmark datasets, including NUS-WIDE \cite{chua2009nus}, MIRFlickr25K \cite{huiskes2008mir} and MS-COCO \cite{lin2014microsoft}, which are widely used in the data retrieval area. In addition, we design a set of ablation experiments to further verify the individual efficacy of different loss components.

\vspace{-2ex}
\subsection{Datasets}
\begin{itemize}
 \item [1)] 
\textbf{NUS-WIDE} contains 269,648 web images and we use the images associated with the 21 most frequent concepts, where each of these concepts associates with at least 5,000 images, resulting in a total of 195,834 images. A total of 2,100 data pairs in the dataset are selected randomly as the query set and the remainder of the dataset is used as the retrieval database. 
 \item [2)]
\textbf{MIRFlickr25K} is a commonly used dataset consisting of 25000 images that were downloaded from the social photography site Flickr.com. In our experiment, we select 20,015 data points in total, among which 10,000 pictures are randomly selected for training. For the remaining data, 2000 data pairs are selected randomly as the query set and the rest is used as the retrieval database.
 \item [3)]
\textbf{MS-COCO 2014} originates from the Microsoft COCO dataset, the 2014 release of MS-COCO contains 82,783 training, 40,504 validation, and 40,775 testing images (approximately 1/2 train, 1/4 val, and 1/4 test). We randomly select 4992 pairs for the query set and leave the remaining pairs as the retrieval database. In addition,  10,000 pairs are randomly selected from the retrieval database for training. 
\end{itemize}

\renewcommand\arraystretch{1.4}
\begin{table}[h]
	\scriptsize
	\centering
	\begin{threeparttable}
	\vspace{-2ex}
		\caption{Experiment settings of different databases.}
		\label{tab:setting}
		\begin{tabular}{c|c|c|c|c}
			\toprule[1pt]
			\multicolumn{2}{c|}{\textbf{Database}}                                          & \textbf{Database Size} & \textbf{Training Size} & \textbf{Category Quantity} \\ \hline
			\hline
			\multirow{2}{*}{\textbf{NUS-WIDE}}     & \textbf{IID}   & 193734 & 10000  & 24
			\\ \cline{2-5} 
			 & \textbf{non-IID}   & 193734 & 10000  & 24
			\\ \cline{1-5} 
			\multirow{2}{*}{\textbf{MIRFlickr}}     & \textbf{IID}   & 18015 & 10000  & 21
			\\ \cline{2-5} 
			 & \textbf{non-IID}   & 18015 & 10000  & 21
			\\ \cline{1-5}
			\multirow{2}{*}{\textbf{MS-COCO}}     & \textbf{IID}   & 112226 & 10000  & 80
			\\ \cline{2-5} 
			 & \textbf{non-IID}   & 112226 & 10000  & 80
			\\ \cline{1-5}
			
			\bottomrule[1pt]
		\end{tabular}
	\end{threeparttable}  
\end{table}
\vspace{-2ex}

\begin{table*}[tp]
\centering
\caption{mAP results for the retrieval task in the IID scenarios.}  
\scalebox{0.8}{
\label{tab:mAP-iid}
\begin{tabular}{|c|c|c|c|c|c|c|c|c|c|}
\hline
\multirow{2}*{Model} & \multicolumn{3}{c|}{NUS-WIDE} &  \multicolumn{3}{c|}{MIRFlickr} & \multicolumn{3}{c|}{MS-COCO}\\ 
\cline{2-10}
        & 12bit(\%) & 24bit(\%) & 48bit(\%) & 12bit(\%) & 24bit(\%) & 48bit(\%) & 12bit(\%) & 24bit(\%) & 48bit(\%) \\
\hline 
DPSH (FedAvg) & 73.78 & 75.14 & 76.29 & 76.92 & 78.71 & 79.32 & 56.48 & 57.86 & 59.54 \\
\cline{1-10}
DPSH (FedProx) & 76.71 & 78.25 & 78.32 & 79.94 & 82.35 & 82.97 & 59.65 & 62.18 & 63.94 \\
\cline{1-10}
DPSH (FedCMR) & 76.87 & 78.21 & 78.64 & 79.45 & 81.36 & 83.41 & 61.41 & 64.13 & 65.75 \\
\cline{1-10}
DPSH (MOON) &77.52  &78.32  &78.89  &78.74  &79.54  &80.32  &56.78  &59.42  &61.73  \\
\cline{1-10}
DCH (FedAvg)  & 70.94 & 73.97 & 75.39 & 76.93 & 79.62 & 80.09 & 55.46 & 57.89 & 58.96 \\
\cline{1-10}
DCH (FedProx)  & 74.28 & 75.98 & 76.53 & 78.97 & 80.90 & 81.42 & 55.54 & 57.31 & 58.99 \\
\cline{1-10}
DCH (FedCMR)  & 74.61 & 76.17 & 76.57 & 79.91 & 81.10 & 81.84 & 56.63 & 58.65 & 59.65 \\
\cline{1-10}
DCH (MOON)  &72.58  &75.76  &77.19  &77.67  &79.92  &81.35  &55.14  &57.80  &60.21  \\
\cline{1-10}
GreedyHash (FedAvg) & 73.56 & 75.83 & 77.38 & 69.37 & 72.64 & 75.82 & 52.65 & 58.79 & 62.64 \\
\cline{1-10}
GreedyHash (FedProx) & 72.45 & 76.17 & 78.02 & 68.72 & 74.86 & 75.05 &50.92 & 55.98 & 61.06 \\
\cline{1-10}
GreedyHash (FedCMR)  & 73.28 & 76.02 & 77.81 & 69.40 & 74.48 & 75.96 & 53.17 & 59.82 & 62.94  \\
\cline{1-10}
GreedyHash (MOON)  &75.87  &77.82  &79.43  &72.56  &76.89  &79.24  &50.83  &55.79  &60.31   \\
\cline{1-10}
CSQ (FedAvg) & 75.61 & 78.02 & 78.94  & 77.56 & 78.53 & 80.85 & 56.86 & 61.13 & 67.31 \\
\cline{1-10}
CSQ (FedProx) & 76.43 & 78.42 & 78.96 & 73.17 & 74.16 & 74.54 & 57.46 & 60.33 & 67.64 \\
\cline{1-10}
CSQ (FedCMR) & 76.71 & 78.74 & 79.05 & 70.56 & 72.29 & 73.91 & 57.05 & 60.57 & 67.26 \\
\cline{1-10}
CSQ (MOON) &73.87  &75.69  &76.91  &71.45  &73.76  &75.43  &57.42  &60.75  &68.54  \\
\cline{1-10}
{\bf FedHAP (ours)}  & {\bf 78.59 } & {\bf 80.31} & {\bf 81.55} & {\bf 86.07} & {\bf 86.17} & {\bf 87.78} & {\bf 66.70} & {\bf 70.14} & {\bf 72.18}\\
\hline
\end{tabular}}
\vspace{1ex}
\end{table*}

\begin{figure*}
  \centering
  	\subfigure[Precision v.s. Recall]{
		\includegraphics[width=0.5\columnwidth]{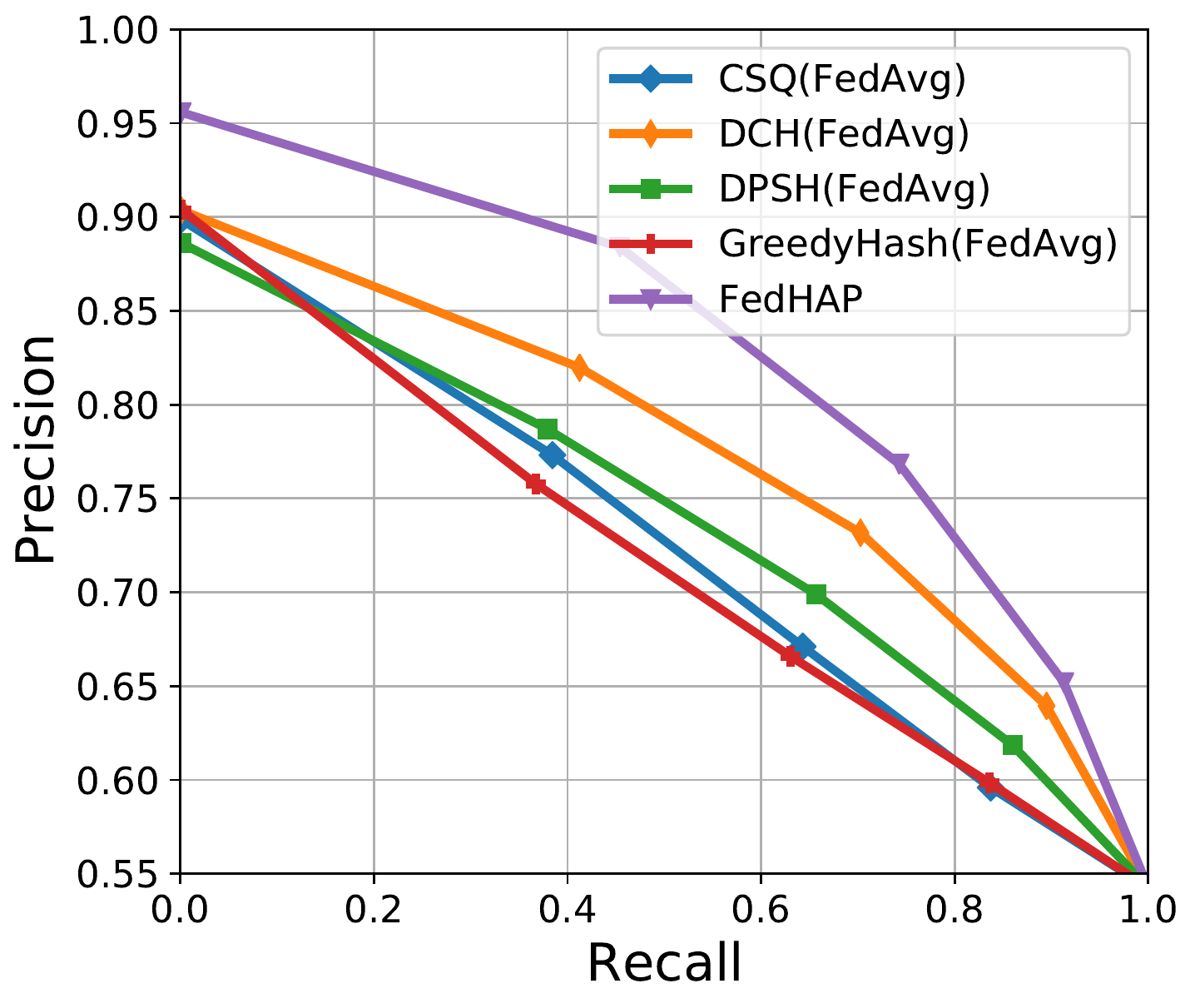}
	}
	\subfigure[Recall v.s. Retrieved samples]{
		\includegraphics[width=0.5\columnwidth]{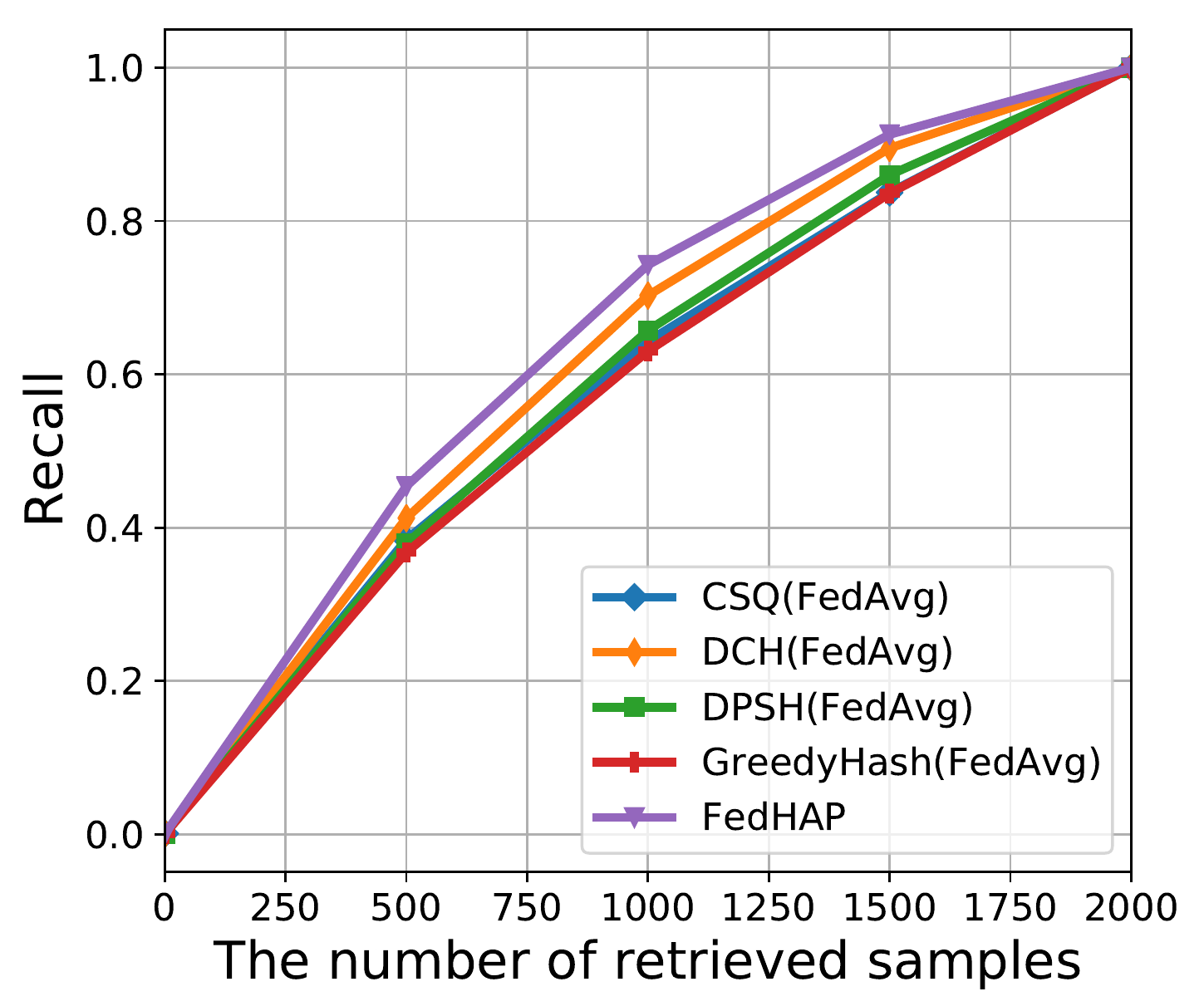}
	}
	\subfigure[Precision v.s. Retrieved samples]{
		\includegraphics[width=0.5\columnwidth]{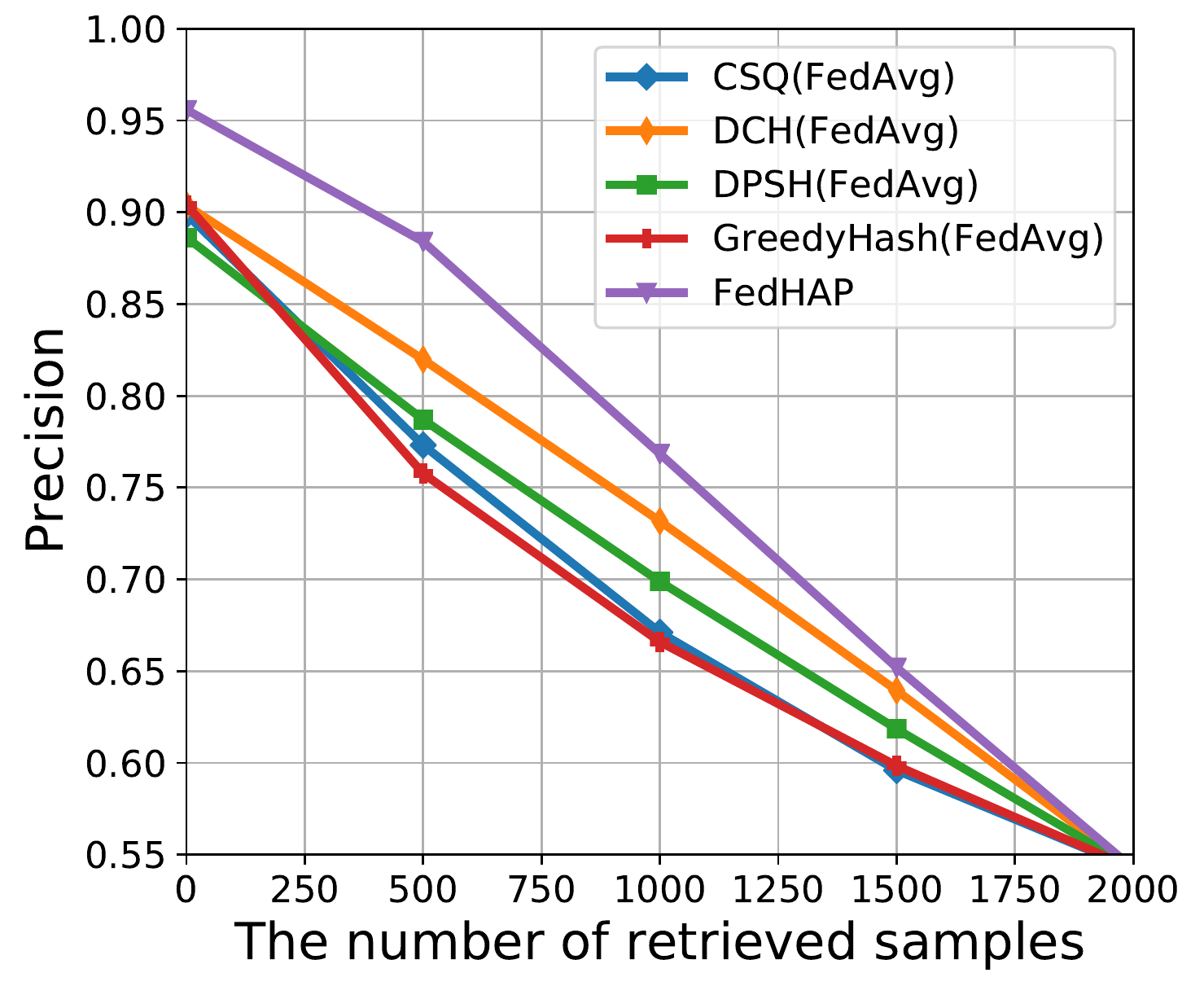}
	}
  \vspace{-2ex}
  \caption{The precision and recall results of DCH (FedAvg), DPSH (FedAvg), Greedyhash (FedAvg), CSQ (FedAvg) and our method FedHAP on MIRFlickr25K in IID scenarios: (a) Precision-recall curves (PR) @ 48 bits. (b) Recall curves with respect to different numbers of top retrieved samples. (c) Precision curves with respect to different numbers of top retrieved samples.}
  
  \label{fig:iid_pr}
\end{figure*}


\begin{table*}[tp]  
\centering
\caption{mAP results for the retrieval task in the non-IID scenarios.}  
\scalebox{0.8}{
\label{tab:mAP-non-IID}
\begin{tabular}{|c|c|c|c|c|c|c|c|c|c|}
\hline
\multirow{2}*{Model} & \multicolumn{3}{c|}{NUS-WIDE} &  \multicolumn{3}{c|}{MIRFlickr} & \multicolumn{3}{c|}{MS-COCO}\\ 
\cline{2-10}
        & 12bit(\%) & 24bit(\%) & 48bit(\%) & 12bit(\%) & 24bit(\%) & 48bit(\%) & 12bit(\%) & 24bit(\%) & 48bit(\%) \\
\hline 
DPSH (FedAvg) & 42.57 & 43.72 & 44.16 & 74.52 & 75.01 & 78.30 & 54.76 & 57.39 & 59.65 \\
\cline{1-10}
DPSH (FedProx) & 44.86 & 45.81 & 45.86 & 74.85 & 75.07 & 78.64 & 56.77 & 59.63 & 60.37 \\
\cline{1-10}
DPSH (FedCMR) & 48.41 & 53.52 & 50.21 & 75.01 & 76.21 & 77.74 & 55.73 & 60.57 & 62.31 \\
\cline{1-10}
DPSH (MOON) &42.67  &47.74  &49.75  &74.87  &76.13  &78.69  &54.65  &57.53  &62.11  \\
\cline{1-10}
DCH (FedAvg) & 38.67 & 39.45 & 40.23 & 70.11 & 70.74 & 77.65 & 51.94 & 56.35 & 58.64  \\
\cline{1-10}
DCH (FedProx)  & 38.42 & 39.25 & 40.98 & 70.81 & 70.81 & 77.82 & 53.18 & 53.28 & 59.25 \\
\cline{1-10}
DCH (FedCMR)  & 40.69 & 43.47 & 44.09 & 69.65 & 69.92 & 73.86 & 53.42 & 57.61 & 58.94 \\
\cline{1-10}
DCH (MOON)  &39.5  &43.21  &44.11  &74.64  &75.08  &77.85  &53.89  &57.80  &58.92  \\
\cline{1-10}
GreedyHash (FedAvg) & 53.68 & 55.82 & 55.94 & 69.93 & 71.21 & 76.60 & 48.57 & 52.16 & 57.25 \\
\cline{1-10}
GreedyHash (FedProx) & 49.29 & 48.41 & 48.90 & 67.59 & 69.62 & 73.56 & 46.57 & 56.80 & 60.41 \\
\cline{1-10}
GreedyHash (FedCMR)  & 51.06 & 63.42 & 63.59 & 65.57 & 69.76 & 70.23  & 48.15 & 55.80 & 55.86 \\
\cline{1-10}
GreedyHash (MOON)  &50.92  &56.79  &61.28  &72.15  &73.46  &76.23   &48.74  &56.51  &57.98  \\
\cline{1-10}
CSQ (FedAvg) & 51.87 & 52.18 & 53.09 & 66.37 & 68.45 & 73.54 & 56.37 & 58.76 & 61.42 \\
\cline{1-10}
CSQ (FedProx) & 54.08 & 55.29 & 55.73 & 69.64 & 72.11 & 73.23 & 50.63 & 58.84 & 62.14 \\
\cline{1-10}
CSQ (FedCMR) & 59.45 & 60.72 & 60.88 & 66.78 & 68.14 & 72.09 & 50.42 & 58.17 & 62.07 \\
\cline{1-10}
CSQ (MOON) &48.76  &53.84  &55.72  &65.52  &68.92  &72.31  &56.71  &58.95  &61.79  \\
\cline{1-10}
{\bf  FedHAP (ours)}  & {\bf 67.74} & {\bf 69.09} & {\bf 70.28} & {\bf 77.34 } & {\bf 78.67 } & {\bf 80.49} & {\bf 57.65 } & {\bf 61.89 } & {\bf 63.37}\\
\hline
\end{tabular}}
\vspace{1ex}
\end{table*}

\subsection{Baselines and Experimental Settings}
We compare FedHAP with the following state-of-the-art deep hashing methods: DCH \cite{cao2018deep}, DPSH \cite{li2015feature}, Greedy Hash \cite{su2018greedy}, and CSQ \cite{yuan2020central}. To ensure a fair comparison with the previous works, the abovementioned methods are deployed under the benchmark learning frameworks: FedAvg \cite{li2020federated}, FedProx \cite{mcmahan2017communication}. We also compare our method with existing federated hashing framework FedCMR \cite{zong2021fedcmr} and MOON \cite{li2021model} which is proposed to handle non-IID local data distributions across clients. The state-of-the-art methods deployed in four federated frameworks are regarded as our baselines and the parameter settings are based on the original papers. 

The feature extraction networks of the baselines are derived from CNN-F \cite{chatfield2014return}, which has been pre-trained on the ImageNet dataset \cite{russakovsky2015imagenet} in order to extract a 4,096-dimensional representation vector for each data point and the discriminator network is a two-layer feed-forward neural network. 

The experiments are conducted in both IID and non-IID scenarios, where the training settings of each scenario are identical for all baselines and our method. We consider a federated learning setup with $m=20$ participating clients. For the IID scenarios, we simulate the IID data distributions by randomly and evenly partitioning the shuffled training sets into 20 clients, and thus each client is assigned with data from a uniform distribution. For the non-IID scenarios, as previous works \cite{mcmahan2017communication, li2020federated}, the data are sorted by class and each client receives a data shard that contains samples belonging to a randomly selected set of classes. It is worth noting that this partition method can result in a deeper heterogeneity of data samples across clients than Dirichlet distribution based partition as in \cite{li2021model}.

In our experiments, the numbers of global training rounds and local training epochs are set to 100 and 5, respectively. In non-IID scenarios, the data category owned by each client is set to 3. Adam \cite{kingma2014adam} is employed as the local optimizer, and the initial learning rate is set to 0.005. The detailed settings of each dataset are summarized in Table.~\ref{tab:setting}. To find a better combination of hyper-parameters in our method, we conduct sensitivity analysis of these hyper-parameters and achieve high results with $\mu=0.05$ and $\lambda=0.1$.

\begin{figure*}
  \centering
    	\subfigure[Precision v.s. Recall]{
		\includegraphics[width=0.5\columnwidth]{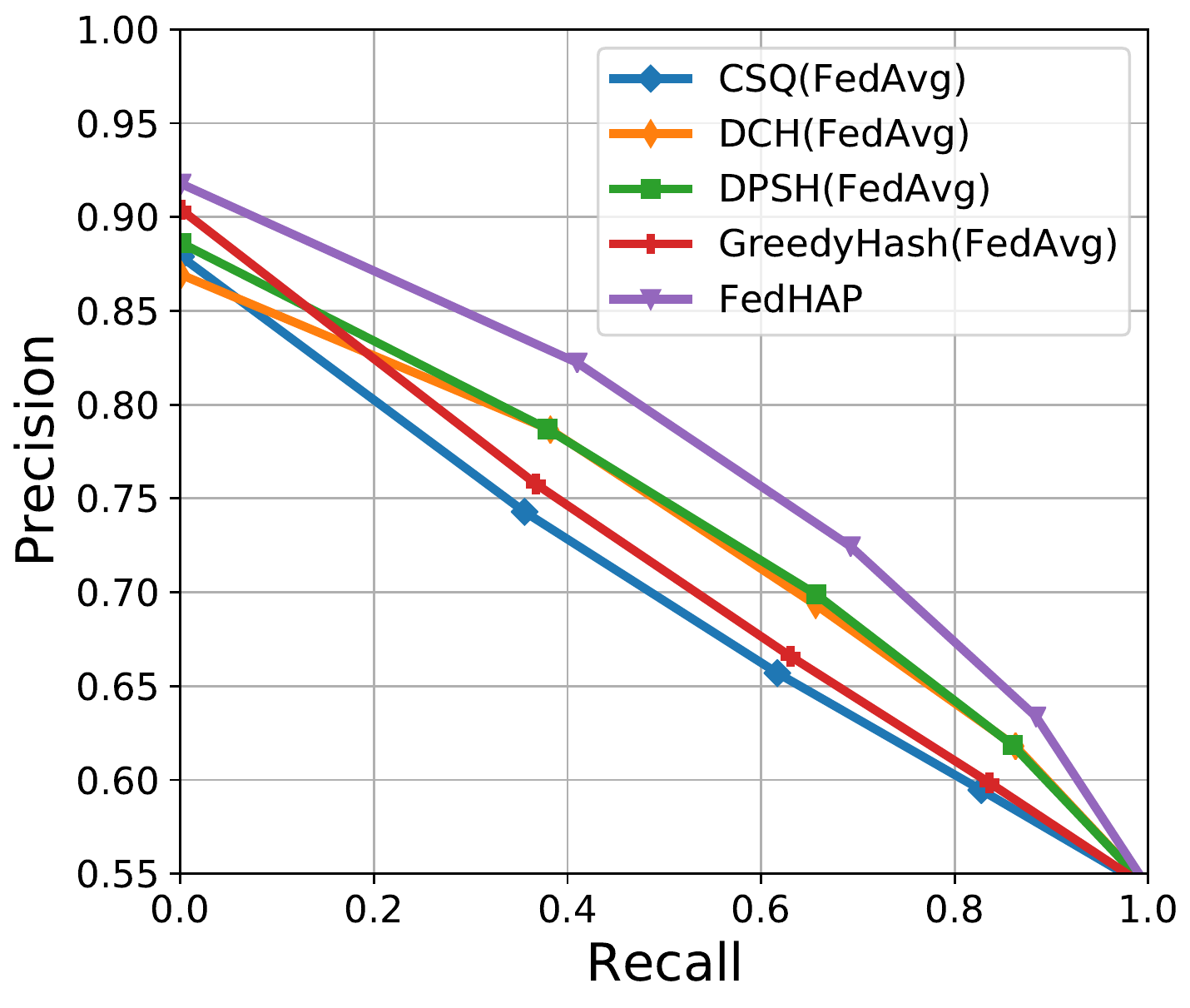}
	}
	\subfigure[Recall v.s. Retrieved samples]{
		\includegraphics[width=0.5\columnwidth]{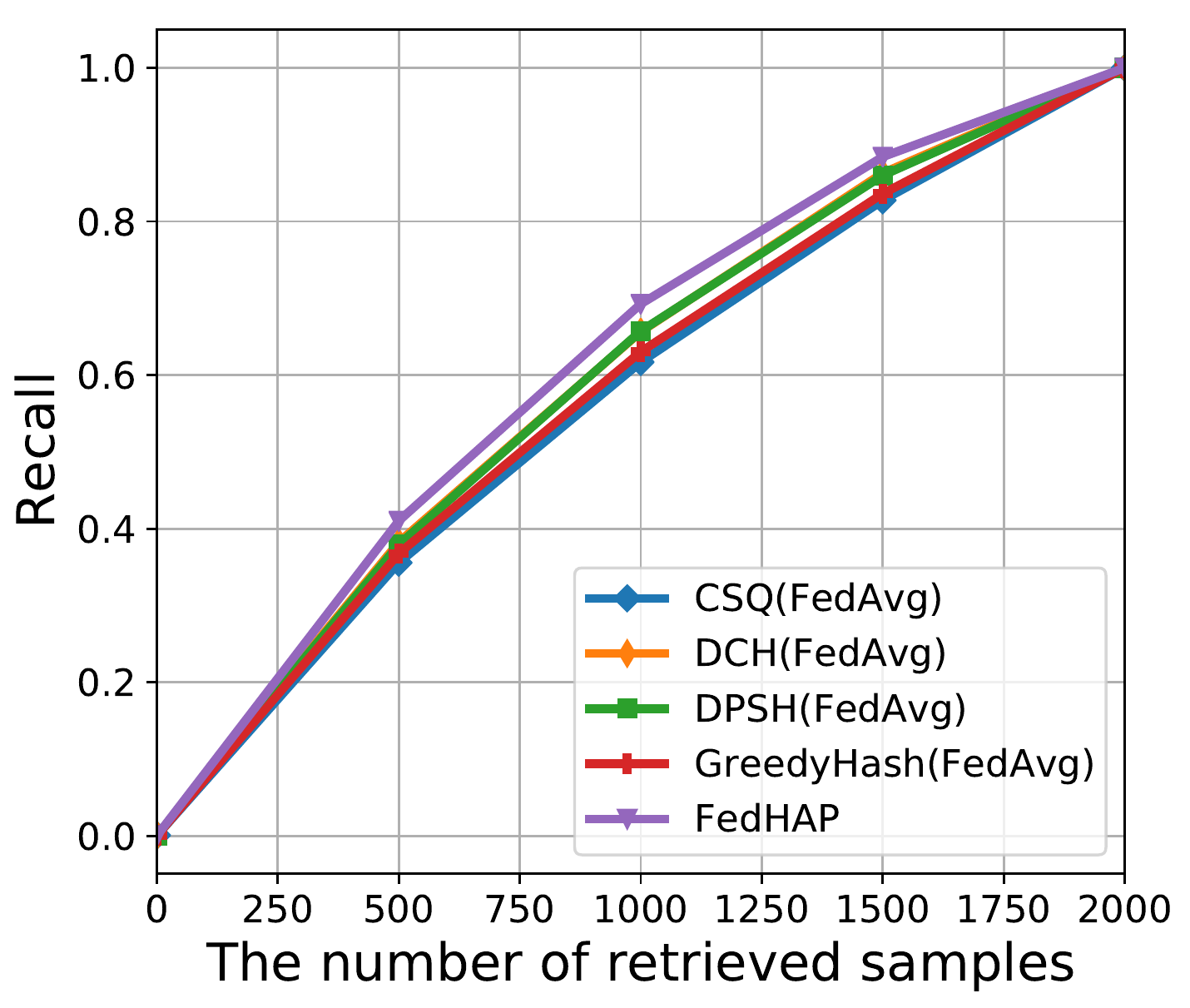}
	}
	\subfigure[Precision v.s. Retrieved samples]{
		\includegraphics[width=0.5\columnwidth]{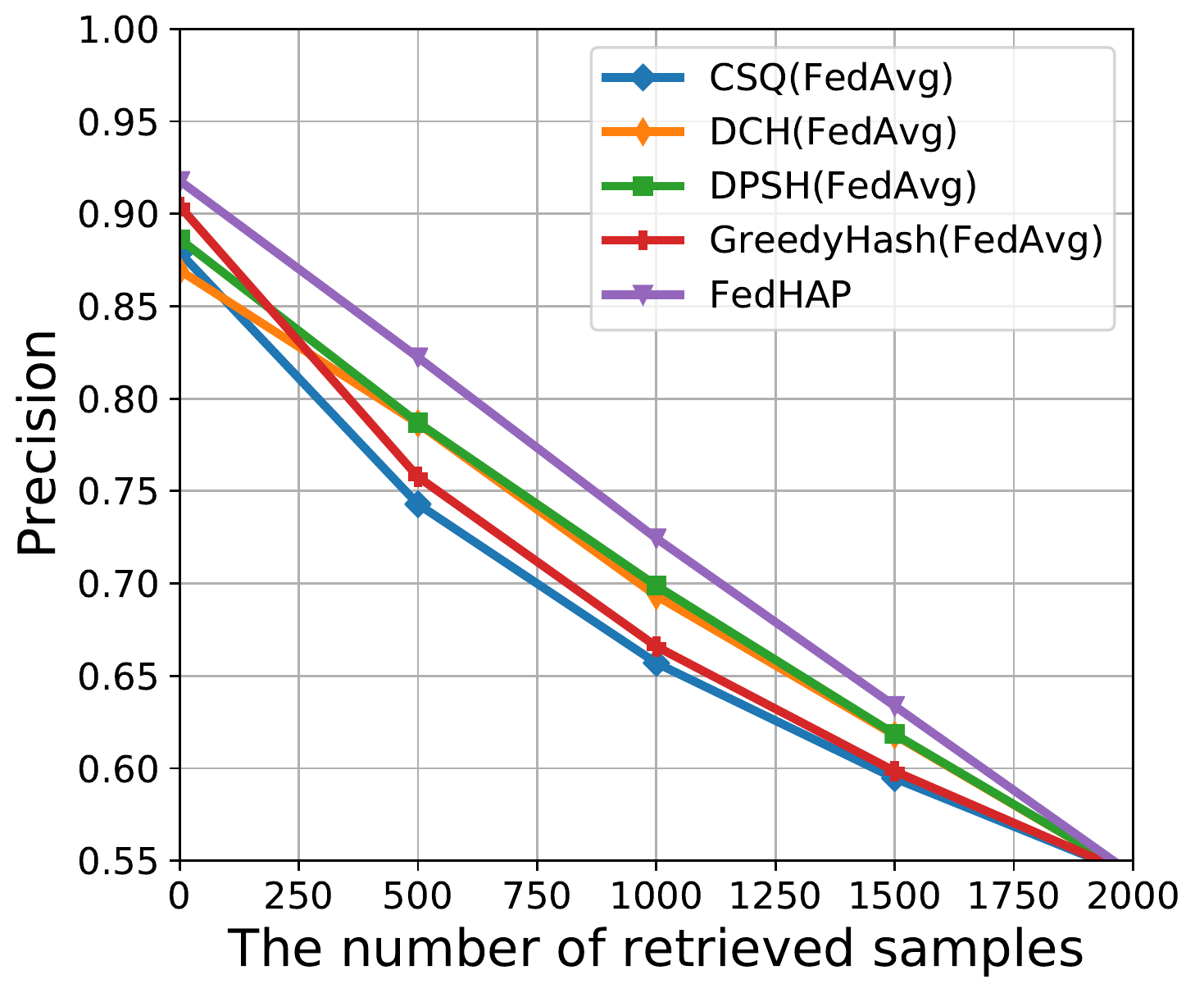}
	}
\vspace{-2ex}
  \caption{The precision and recall results of DCH (FedAvg), DPSH (FedAvg), Greedyhash (FedAvg), CSQ (FedAvg) and our method FedHAP on MIRFlickr25K in non-IID scenarios: (a) Precision-recall curves (PR) @ 48 bits. (b) Recall curves with respect to different numbers of top retrieved samples. (c) Precision curves with respect to different numbers of top retrieved samples.}
  \label{fig:noniid_pr}
\end{figure*}

\subsection{Evaluation Metric}
Hamming ranking is a kind of classical retrieval method that is used to evaluate the performance of the image retrieval task. In our experiments, we evaluate the retrieval quality based on Mean average precision (mAP). As an intuitive illustration, the standard evaluation metrics, including precision-recall curves (PR), recall curves with different numbers of top retrieved samples and precision curves with different numbers of top retrieved samples on MIRFlickr25K dataset, are also provided. For a fair comparison, all methods use the identical training and testing sets.

\vspace{-2ex}
\subsection{Performance Comparison}
We validate the effectiveness of data retrieval in federated environments as well as the generality of our approach in different databases. Our approach is compared with the baselines mentioned above using mAP results, precision-recall curves (PR), precision curves and recall curves, which are shown in Table.~\ref{tab:mAP-iid}, Table.~\ref{tab:mAP-non-IID}, Fig.\ref{fig:iid_pr} and Fig.~\ref{fig:noniid_pr}. In terms of the mAP results, it can be found that regardless of the IID or non-IID scenarios, our approach achieves the best performance in all three databases. 

\noindent \textbf{Results in the IID scenarios.}
Compared with the existing methods, our method further improves the mAP results by approximately 1-2\%, 4-6\% and 5-7\% under the constraints of hash codes with different bits on NUS-WIDE, MIRFlickr25K and MS-COCO, respectively. Moreover, it is noticeable that the improvement of mAP on MS-COCO is much larger than that on the other two datasets. Considering that MS-COCO has the largest data categories and thus the average amount of per-class samples at each client is much less, it can be concluded that our proposed method can overcome the scarcity of local data by leveraging global prototypes and reduce the local over-fitting risks, thus achieving improved model performance.

\noindent  \textbf{Results in the non-IID scenarios.}
In the non-IID scenarios, we also achieve significant improvements of 8-9\%, 2-4\% and 1-2\% in average mAPs for different bits on the above three datasets, respectively. An interesting phenomenon is that the performance boost on MS-COCO in the non-IID scenario is slightly reduced. This may be caused by the fact that the total number of data categories in MS-COCO exceeds the total number of data categories of the other two datasets, which will lead to an increased non-IID degree across clients. 

The extensive retrieval performance results on MIRFlickr25K with regard to precision-recall curves (PR), precision curves and recall curves with respect to different numbers of top returned samples in Fig.\ref{fig:iid_pr} and Fig.~\ref{fig:noniid_pr} show that FedHAP outperforms baseline methods impressively, which is desirable for practical precision-first retrieval. Specifically, FedHAP achieves higher precision when the recall levels are low or the number of retrieved samples is small. In conclusion, these results demonstrate that learning the hash function using our proposed method can boost the retrieval performance remarkably in federated environments.

\subsection{Analysis on Ablation Experiments}
To better demonstrate our contributions, we design a set of ablation experiments to verify the utility of different components in our FedHAP framework. The ablation experiment is defined as:

\textbf{FedHAP-1}: 
FedHAP-1 is designed based on FedHAP without the participation of global prototypes, which means that the process of discriminator network training and the global prototypes’ participation in similarity preserving loss calculation are all removed. The remainder of the method is the same as FedHAP.

\textbf{FedHAP-2}: 
FedHAP-2 is built based on the design of FedHAP, in which the global prototypes do not participate in the calculation of similarity preserving loss but still participates in the training process of the discrimination network, which means we only apply the global prototypes to the adversarial module. 

\textbf{FedHAP-3}: FedHAP-3 is designed based on FedHAP. Contrary to FedHAP-2, global prototypes only participate in the calculation of similarity preserving loss and the module of the discriminator network is removed in the framework. 

\renewcommand\arraystretch{1.2}
\begin{table}[ht]
\centering
  \setlength{\tabcolsep}{0.8mm}
  \caption{mAP results of the ablation experiments (IID).}
  \vspace{-0.2cm}
  \begin{tabular}{cccccccl}
    \toprule
    \multirow{2}*{Method} & \multicolumn{3}{c}{NUS-WIDE} & \multicolumn{3}{c}{MS-COCO}\\
    \cmidrule(lr){2-4}\cmidrule(lr){5-7}
    & 12bit(\%) & 24bit(\%) & 48bit(\%) & 12bit(\%) & 24bit(\%) & 48bit(\%)\\
    \midrule
    FedHAP-1  & 74.67 & 75.83 & 76.44 & 62.23 & 63.77 & 68.37\\
    FedHAP-2  & 75.43 & 77.93  & 78.78 & 65.42 & 68.09 & 71.71\\
    FedHAP-3  & 76.74 & 78.09 & 80.02 & 65.63 & 68.29 & 71.19\\
    FedHAP  & {\bf 78.59 } & {\bf 80.31} & {\bf 81.55} & {\bf 66.70 } & {\bf 70.14 } & {\bf 72.18}\\
    
  \bottomrule
\label{tab:ab-iid}
\end{tabular}
\vspace{-2ex}
\end{table} 

\vspace{-4ex}

\renewcommand\arraystretch{1.2}
\begin{table}[ht]
 \centering
   \setlength{\tabcolsep}{0.8mm}{}
  \caption{mAP results of the ablation experiments (non-IID).}
  \vspace{-0.2cm}
  \begin{tabular}{cccccccl}
    \toprule
    \multirow{2}*{Method} & \multicolumn{3}{c}{NUS-WIDE} & \multicolumn{3}{c}{MS-COCO}\\
    \cmidrule(lr){2-4}\cmidrule(lr){5-7}
    & 12bit(\%) & 24bit(\%) & 48bit(\%) & 12bit(\%) & 24bit(\%) & 48bit(\%)\\
    \midrule
   FedHAP-1 & 57.16 & 59.66 & 60.56 & 55.60 & 58.59 & 60.85 \\
    FedHAP-2 & 64.27 & 65.60 & 66.51 & 56.31 & 60.22 & 61.89 \\
    FedHAP-3 & 65.61 & 67.99 & 68.18 & 56.80 & 60.63 & 61.86 \\
    FedHAP  & {\bf 67.74} & {\bf 69.99} & {\bf 70.28} & {\bf 57.65} & {\bf 61.89} & {\bf 63.37}\\
    
  \bottomrule
\end{tabular}
\label{tab:ab-non-IID}
\end{table}

\vspace{-2ex}

The results of ablation experiments in IID and non-IID scenarios are reported in Table.~\ref{tab:ab-iid} and Table.~\ref{tab:ab-non-IID}. Two points can be concluded from the results. First, comparing the results of FedHAP-1 and FedHAP, it can be found that the model performance will degrade significantly in the absence of global prototypes, which demonstrates the efficacy and importance of global prototypes in promoting retrieval performance. Second, each component in the framework can play a significant role in improving the model performance independently. The optimal results are achieved through the mutual promotion between different components in the FedHAP framework.



\subsection{Effect of The Number of Clients}
To analyze the performance of our proposed method when the client number varies, we further test the above-mentioned baselines and our method with different numbers of clients from 20 to 100, where the data samples are randomly distributed and the length of hash code is 48 bits. The mAP results are reported in Table~\ref{tab:silos}, from which we can see that our method still consistently outperforms all baselines under different system sizes. We also notice that as the number of clients increases, the performance of the model will decrease slightly. This is not surprising since the amount of data per client will decrease when the number of clients increases and the total amount of data remains the same, which will result in enlarged distribution discrepancy of local data and higher probability of local model over-fitting.
\renewcommand\arraystretch{1.5}
\begin{table}[h]
	\scriptsize
	\centering
	\setlength{\tabcolsep}{2.6mm}{}
	\begin{threeparttable}
		\caption{mAP results  under different numbers of clients.}
		\vspace{-2ex}
		\label{tab:silos}
		\begin{tabular}{c|c|c|c|c|c}
			\toprule[1pt]
			\multicolumn{2}{c|}{\textbf{Method}}                                          & \textbf{20 clients} & \textbf{40 clients} & \textbf{60 clients}& \textbf{100 clients} \\ \hline
			\hline
			\multirow{2}{*}{\textbf{DPSH}}     & \textbf{FedAvg}   &79.32  &79.18   &78.93  &78.42
			\\ \cline{2-6} 
			 & \textbf{FedProx}   &82.97  &82.80  &82.54 &82.21
			\\ \cline{1-6} 
			\multirow{2}{*}{\textbf{CSQ}}     & \textbf{FedAvg}   &80.85  &79.81   &79.46  &79.10
			\\ \cline{2-6} 
			 & \textbf{FedProx}   &74.54  &73.28   &72.91 &72.65
			\\ \cline{1-6} 
			\multicolumn{2}{c|}{\textbf{Our method}}     &\textbf{87.78}    &\textbf{87.25}  &\textbf{87.16}   &\textbf{86.83} 
			\\ \cline{1-6}
			
			\bottomrule[1pt]
		\end{tabular}
	\end{threeparttable}  
\end{table}

\subsection{Effect of Distance Metrics}
Here, we compare the results of our FedHAP using two different distance metrics in computing triplet loss, including the Euclidean distance and cosine distance, reporting the results in Table~\ref{tab:distance-non-IID}. We can observe that both Euclidean distance and cosine distance can significantly improve the performance compared to the baselines in Table~\ref{tab:mAP-non-IID}, and the cosine distance outperforms the Euclidean distance. We consider the reason is that the cosine distance could eliminate the influence of different norms of output feature vectors. 

\begin{table}[h]
 \centering
  \setlength{\tabcolsep}{2mm}{}
  \caption{ mAP results under different distance metrics in non-IID scenarios over 48 bits.}
  \vspace{-0.2cm}
  \begin{tabular}{ccccl}
    \toprule
    Distance Metric & NUS-WIDE & MIRFlickr & MS-COCO \\
    \midrule
    Euclidean Distance & 69.41 & 79.23 & 62.78  \\
    Cosine Distance  & {\bf 70.28} & {\bf 80.49} & {\bf 63.37} \\

  \bottomrule
\end{tabular}
 \vspace{-0.3cm}
\label{tab:distance-non-IID}

\end{table}

\section{Conclusion}
In this paper, we propose a novel federated hashing approach FedHAP for efficient cross-silo retrieval, which aims to collectively train the hashing models from decentralized data. Besides the general federated manner, we innovatively introduce the global prototypes to maintain the distribution alignment of the locally generated and globally generated hash codes, achieving a significant improvement in the model effectiveness. Since the global prototypes are composed of fixed-length (12-48 bits) binary hash codes and the number of the hash codes does not exceed the number of data categories, which guarantees almost negligible communication cost and does not raise data privacy issues. Comprehensive experimental results on three widely used databases have demonstrated the superiority of FedHAP compared with other baselines in both IID and non-IID scenarios. 

\bibliographystyle{ACM-Reference-Format}
\bibliography{sample-base}

\clearpage


\end{document}